# 3D core@multishell piezoelectric nanogenerators

A. Nicolas Filippin,[a] Juan R. Sanchez-Valencia,[a,b] Xabier Garcia-Casas,[a] Victor Lopez-Flores,[a] Manuel Macias-Montero,[a] Fabian Frutos,[c] Angel Barranco[a] and Ana Borras[a]*

[a.] Nanotechnology on Surfaces Group, Materials Science Institute of Seville (CSIC-University of Seville), C/ Américo Vespucio 49, 41092, Seville, Spain.

[b.] Departamento de Física Atómica, Molecular y Nuclear, Universidad de Sevilla, Avda. Reina Mercedes s/n, 41012, Seville, Spain.

[c.] Departamento de Física Aplicada I, Universidad de Sevilla, Avda. Reina Mercedes s/n, 41012, Seville, Spain.

*Corresponding author e-mail: anaisabel.borras@icmse.csic.es

**Abstract**

The thin film configuration presents obvious practical advantages over the 1D implementation in energy harvesting systems such as easily manufacturing and processing and long lasting and stable devices. However, most of the ZnO-based piezoelectric nanogenerators (PENGs) reported so far relay in the exploitation of single-crystalline ZnO nanowires because their self-orientation in the c-axis and ability to accommodate long deformations resulting in a high piezoelectric performance. Herein, we show an innovative approach aiming to produce PENGs by combining polycrystalline ZnO layers fabricated at room temperature by plasma assisted deposition with supported small-molecule organic nanowires (ONWs) acting as 1D scaffold. The resulting hybrid nanostructure is formed by a single-crystalline organic nanowire conformally surrounded by a three dimensional (3D) ZnO shell that combines the mechanical properties of the organic core with the piezoelectric response of the ZnO layer. In a loop forward towards the integration of multiple functions within a single wire, we have also developed ONW@Au@ZnO nanowires including a gold shell acting as inner nanoscopic electrode. Thus, we have built and compare thin films and 3D core@shell ONW@ZnO and ONW@Au@ZnO PENGs showing output piezo-voltages up to 170 mV. The synergistic combination of functionalities in the ONW@Au@ZnO devices promotes an enhanced performance generating piezo-currents almost twenty times larger than the ONW@ZnO nanowires and superior to the thin film nanogenerators for equivalent and higher thicknesses.

# 1.- Introduction.

Low-cost/high-throughput self-powered nanosystems are called to revolutionize remote monitoring, healthcare, wearable electronics and communications. These devices operate by sensing, controlling, communicating, and actuating/responding thanks to integrated power nanogenerators (NGs) able to harvest and convert energy from the local environment.[1-2] Current approaches in this domain exploit photovoltaic, thermoelectric, chemical processes triboelectric or piezoelectric effects to generate electricity. These two latter aim to convert into electricity kinetical energy harvested from low frequency movements (body movement, muscle stretching), vibrations (acoustic or ultrasonic waves) and hydraulic sources (such as body fluid flow and ocean waves).[3] Single-crystalline ZnO NWs are ubiquitous components in piezoelectric nanogenerators (PENGs),[4-6] meanwhile, other materials like ZnS, GaN, CdS, $BaTiO_3$ NWs and PZT nanofibers are attracting more attention during the last years.[3,7] Nanofibers made of polyvinylidene fluoride (PDVF) represent the organic counterpart in the field with a high performance in both single-wire and bundle operation.[8] As first step for a competitive realization of the self-powered nanosystems, high power output and long term stable NGs are required. Several approaches have been proposed so far including the development of different piezo-nanomaterials and the optimization of ZnO-Piezo NGs. The NG classic VING (vertically integrated nanogenerator) architecture[9] with the single crystalline NWs in intimate contact between a Schottky and ohmic electrode has evolved during the last years in order to reduce screening effects at the interface with contacts and to increase the piezo-output voltage.[10] Reference [11], gathers different examples of the combination of ZnO NWs with insulating (poly(methyl methacrylate, PMMA) and p-type semiconducting polymers (poly(3,4-ethylenedioxythiophene) polystyrene sulfonate PEDOT:PSS). Additional improvements in NGs including p-n junctions have been obtained by coating the ZnO surfaces with a thin layer of p-type CuSCN and using a more conductive p-type polymer (p-type polymer poly(3,4-ethylenedioxythiophene-Tosylate, PEDOT:Tos).[12] Although the voltage generated in these devices is still lower than those of devices containing PMMA–Au Schottky contacts, the current density is much higher due to the low internal resistance meaning that an overall relatively high level of power can be transferred to a load. Historically, the first examples of nanogenerators were developed using ZnO NWs. These ZnO nanomaterials are still indispensable for the development of NGs and piezotronic devices. Nevertheless, they suffer from several limitations that have discouraged their industrial adoption such as a lack of uniformity in the synthetic methodologies, difficult circuit integration and large performance variations in single-crystal NWs devices. [1, 13] At present, the research is mainly focussed in thin film piezoelectric devices granted by self-oriented growth of the ZnO grains along the c-axis.[14]

Within that context, herein we present the unprecedented application of core@multishell supported nanowires as piezoelectric nanogenerators. To the best of our knowledge, this is the first time that such approach is intended including the use of supported single-crystalline organic nanowires (ONWs) as 1D template [15-19] and combining polycrystalline ZnO and Au shells to form 3D core@multishell nanoarchitectures.[20-21] These core@multishells are formed by a polycrystalline ZnO layer acting as active piezoelectric shell and an Au layer intimately connecting the nanowire with the conducting substrate. With such approach we aim to combine the advantages resulting from the synthesis of ZnO layers by plasma enhanced chemical vapour deposition (PECVD) with the advantages of the 1D and 3D nanomaterials implementation. PECVD ZnO nanostructured film synthesis present high homogeneity and reproducibility, low temperature and solvent-less operation, ample margin in crystallinity, composition, and microstructure optimization and compatibility with an impressive variety of substrates.[22-23]. The implantation in 1D-3D architectures permits to develop high-density arrays with enhanced response to mechanical stimulus. In addition, the possibility of adding different layers to the core@multishell architecture open the path towards a new generation of highly integrated and multifunctional piezoelectric and piezotronic nanowire devices.

In this communication, we will first present the vacuum and plasma deposition methodology for the formation of core@multishell nanowires and then, in the second part, promising results comparing thin films, polycrystalline ZnO NWs and multifunctional Au-ZnO NWs piezoelectric nanogenerators.

## 2.- Experimental

Organic NWs where growth by vacuum deposition using a LTE01evaporation source from Kurt J. Lesker. The base pressure of the chamber was lower than $10^{-6}$ mbar. The total pressure during deposition was 0.02 mbar of Ar, the deposition rate was 0.45 Å/s (measured by QCM) and the substrate temperature was 210 ºC. The sample-to-evaporator distance was 6.5 cm.

The deposition of a conformal ZnO was carried out by PECVD at room temperature. Diethylzinc (($CH_2H_5)_2Zn$) from Sigma-Aldrich was used as delivered as Zn precursor. The precursor and oxygen were delivered in the plasma reactor by using mass flow controlers. The plasma was generated in a 2.45 GHz microwave Electron-Cyclotron Resonance (ECR) operating at 400 W. The deposition was carried out in the down-stream region of the reactor. The film thicknesses were controlled with a calibrated quartz crystal monitor (QCM).

The deposition of metallic seeds for the ONWs deposition was carried out by magnetron sputtering at 0.1 mbar employing and Emitech K550 sputter coater equipped with a gold target. 12.5 mA 15 s yielded adequate Au NPs both in density and size. The deposition of the Au shells was carried out in the same experimental set-up using and 25mA 45s to achieved full percolation.

For the fabrication of some piezoelectric devices, approximately 500 nm of PMMA were spin coated (5 wt/v % in acetone) on selected core@shells samples followed by heating in air at 80 ºC for 1 hour to facilitate the polymer infiltration. Spin coating was performed in a WS-400-6NPP-LITE coater from Laurell.

Room temperature plasma etching of PMMA was carried out with the samples facing the plasma during 15 minutes at 400 W and 0.02 mbar of oxygen as plasma gas.

SEM micrographs were acquired in a Hitachi S4800 working at 2 kV, while STEM (SEM) micrographs were acquired in a Hitachi S5200 working at 30 kV. The samples were dispersed onto Holey carbon films on Cu or Ni grids from Agar scientific for TEM characterization. HAADF STEM and HRTEM were carried out with both FEI Tecnai Orisis TEM/STEM 80-200 and FEI Tecnai G2F30 S-Twin STEM microscope also working at 200 kV.

The crystal structure was analyzed by XRD in a Panalytical X'PERT PRO spectrometer operated in the θ - 2θ configuration and using the Cu Kα (1.5418 Å) radiation as an excitation source.

I-time and I-V curves were recorded with a Keithley 2635A system sourcemeter working in open circuit conditions (V = 0) and sweep voltage mode correspondently.

## 3.- Results and Discussion

Scheme 1 presents the synthetic methodology applied in the formation of the supported core@multishell nanowires. This protocol allows the fabrication of hybrid nanowires formed by organic single crystalline core (ONW) acting as well as 1D scaffold, conformally surrounded by inorganic shells. The organic core is removed providing supported open or domed nanotubes with a simple annealing in vacuum or air.[20-21] In recent articles, we have successfully applied such approach for the alignment of magnetic molecular wires,[18] formation of nanoscale waveguides,[19] development of 1D and 3D photoanodes for dye-sensitized solar cells[24] and the fabrication of tuneable superhydrophobic-superhydrophilic surfaces.[25] In this communication, we aim to do a loop forward towards the exploitation of the advantages of the method by demonstrating its successful application to develop highly integrated energy harvesting devices. Step 0) in the scheme consists in the formation of seeds or nucleation centres on the desired substrate to promote the crystallization process responsible for the ONWs formation. The chemical nature of such centres is not relevant as we have reported elsewhere [16-17,19]. Thus, metal nanoparticles, metal or metal oxide thin films or polymers may act as support for the growth of the nanowires. Looking to the final application of the core@shell NWs as PENGs, we use herein Si(100) and fused silica reference substrates as well as commercially available ITO/PET supports foils decorated with a ZnO thin film by plasma enhanced chemical vapour deposition (PECVD).[21-23] Figure 1 a-b) shows typical cross section and normal view SEM micrographs of the ZnO thin films. The microstructure, texture and photoluminescence properties of the thin films can be tuned controlling experimental parameters as plasma gas composition, substrate temperature and growth rate.[21,23] The deposition under room temperature conditions, oxygen as plasma gas and low growth rate (see Methods an Supporting Information sections for experimental details) produces columnar thin films with granular surfaces presenting preferential formation of (101) planes, as shown in the XRD pattern in Figure 1 c). The photoluminescence spectrum corresponding to such film (Figure S1 in the Supporting Information) is dominated by the exciton peak at 380 nm with a broad low intense visible emission in the region 450–700 nm. The ZnO visible emission have been attributed to defects as zinc vacancies (blue emission) and oxygen vacancies (green emission).[ 21-23] For ZnO PECVD films, the level of such defects is low, as addressed by the low intensity of the visible emission what have been attributed to a suppression of band-gap electronic defects by hydrogen species from the plasma during the synthesis.[22]

The ONWs formation is achieved during Step I) by vapour transport of small pi-conjugated molecules (porphyrins, phthalocyanines or perylenes) and self-assembly into single-crystalline nanowires. ONWs form in either squared morphologies or nanobelts with thicknesses in the range between 50 and 200 nm, depending on the size of the nucleation centres, and lengths from 0.5 to several tens of micrometres. The density of nanowires and their lengths can be controlled through the deposition time (from minutes up to a couple of hours).

We have optimized our system for the formation of high density arrays of (H2- and metal)-phthalocyanine and (H2- and metal)-porphyrin nanowires at temperatures ranging from 100 to 220 ºC. Note that this procedure is compatible with the use of organic nanowires formed at room temperature, as reported for anthraquinone and other small molecules.[26] The examples gathered in this communication correspond to H2-phthalocynanine (H2Pc) nanowires.

During Step II) the ONWs are conformally cover in order to form hybrid core@shell NWs. We have included two different inorganic shells. In the first place, an Au gold layer was deposited by magnetron sputtering. The thickness of such layer was settled in order to ensure the electrical connectivity and flexibility of the wires to accommodate both, the second inorganic layer and the mechanical deformation. The thickness for the equivalent Au thin film, i.e. deposited during the same experiment on a flat substrate (Figure S2), is ca. 70 nm. For this layer, we have estimate a sheet resistivity of ~ 8 Ω/□ by four-point probe measurement. Figure S2 details the morphology of the Au layer forming the H2Pc@Au nanowires that still appear bended and twisted in randomly directions. It is worthy stressing that, although we have selected Au as inner electrode in order to build the final devices complying with the ohmic-Schottky required combination, this protocol is compatible with the deposition of other metallic or transparent conducting oxide shells acting as inner electrode. Among them, we can cite thermal or electron beam evaporation and atomic layer deposition (ALD) granted that the operation temperature is maintained below the sublimation temperature of the molecules (usually below 320 ºC for phthalocyanines). In the case of using the ONWs as 1D templates for nanotubes synthesis the thermal restriction have been solved by depositing a conformal initial shell at mild temperatures acting as a 1D scaffold to sustain subsequent higher deposition temperatures. [24] Finally, for the formation of the piezoelectric ZnO shell we have applied the same PECVD process than during Step 0). One of the major advantages of this method is that working under a down-stream configuration allows the deposition and vertical alignment of the NWs without damaging their crystalline structure.[19] Figure 1 d-f) gathers the cross section and planar view images of the resulting array of $H_2Pc@Au@ZnO$ nanowires. The external morphology does not differ from the $H_2Pc@ZnO$ nanowires formed without the inner Au shell. Thus, the nanowires appear vertically aligned with respect to the substrate with a thicker tip consequence of self-shadowing effects during the growth of the inorganic shell. In this case, for an equivalent thickness of the ZnO layer of 400 nm, the @ZnO NWs (addressing the ZnO shells at both the H2Pc@ZnO and H2Pc@Au@ZnO nanowires) present a medium thickness of 200 nm with a minimum coverage of ~ 20 nm estimated at the interface with the substrate. The nanowire density reaches up to 10 $NW/um^2$. It is interesting to indicate the density values and the mean nanowire diameter are parameters can be controlled through the experimental conditions. Hence, we expect in forthcoming experiments to fine-tune these parameters in order to optimize the final output power of the NGs based in such core@shell structures. Figure 1 g-h) presents a detail on the inner microstructure of an individual $H_2Pc@Au@ZnO$ nanowire, where the 10 nm Au and ZnO shells surrounding the squared organic nanowire are clearly visible. The ZnO layer forms a 3D shell of small columns radially distributed around the nanowire. In consequence, the XRD pattern corresponding to the ZnO shell (Figure 1c) presents a slight texturization in the (100) direction. indicating the radially distributed nanocolumns preferentially grow with the (100) planes parallel to the substrate. This result agrees with the HRSTEM characterization of the @ZnO shells in reference [20] elucidating the presence of both (100) and (002) planes in the ZnO grains. These polycrystalline core@shell nanowires aim to compete with single-crystalline ZnO nanowires. Besides, it is interesting to call the attention towards the low level of defects appearing in the @ZnO nanowires even though we have yet not achieved a fully texturized surface. For simplicity, we have carried out their photoluminescence (PL) characterization after evacuation of the organic core, otherwise also luminescent. As in the case of the ZnO thin films, the PL spectrum (Figure S1) is fully dominated by the exciton emission in the UV range, showing a defect related visible emission of even lower intensity than for the reference film. This result indicates that the number of defects in the conformal ZnO shell does not increase in comparison with the thin film counterpart although both PL spectra are not entirely equivalent. The PL exciton band of the ZnO nanotubes is broader and red-shifted what is congruent with a higher stress, smaller crystallite size and lower texture development of the ZnO shells. [22-23]Moreover, a certain additional contribution to the observed red-shift of the UV PL band due to the anisotropic character of the oriented ZnO nanowires emission cannot be discarded (see [22] and references cited therein).

We follow now to present the integration of core@multishell nanowires as active part in piezoelectric nanogenerators. As previously mentioned, most of the examples in literature of ZnO piezoelectric nanogenerators develops around the mechanical deformation of single-crystalline ZnO nanowires along the c-axis (planes (001)). To optimize the output power generation and simplify the device operation, the ZnO nanowires very often appear perpendicular to the substrates and sandwiched between two different electrodes, usually combining electrodes acting with the piezoelectric materials as ohmic and Schottky contacts. In a first place and for the sake of simplicity, it is interesting to prove how the PECVD polycrystalline ZnO thin films convert mechanical deformations into electrical current. Figure 2 gathers characteristic I-V, I-t and V-t curves of the studied different systems. For the thin film NGs, we have deposited the ZnO layers on commercially available ITO/PET substrates and contacted the ZnO and ITO with cooper tape. In order to improve the lifetime of the devices and to facilitate the mechanical deformation, the entire system was embedded between two PDMS foils with about 2 mm of thickness (see Methods). Note that, the difficulties to estimate the actual active surface for the NWs may hampers the straightforward comparison between the thin film and NW devices (see a later discussion over this point). Figure 2 a) shows the characteristic I-V curve for a 900 nm ZnO thin film comparing un-loaded and loaded scenarios. The I-V curve depicts the expected rectifying behaviour with no evidence of short circuits. Under load condition, the curve presents an increase in current for the same voltage values than for the unloaded case. In fact, these results constitutes the first evidence of piezoelectric effect in the film. The shift in the I-V curves for the loaded and unloaded conditions allows the estimation of the piezo-voltage output for a given current.[27-28] Thus, for a current of 15 nA, corresponding to the maximum current obtained with this device, the estimated voltage output is 105 mV. This a relatively high voltage value, in line with those reported for single-crystalline ZnO nanowires. [29-30] The response to intermittent pressure of this system is presented in Figure b-c). From these curves is evident that the device generates current as consequence of the applied deformation in AC configuration. In this arrangement, the negative peak corresponds to the downside deformation and the positive spike to the relaxation of the system. It is worthy to mention that the change in polarity of the contacts reverse the AC behaviour, as expected in this type of NGs.[7] In order to evaluate the role of the thickness of the thin film, we prepared a second device with a thicker ZnO layer (~2.4 µm) deposited under the same conditions but during a longer time. Panels d and e) present the corresponding results. For equivalent maximum load, the thicker thin film presents a higher output current of 70 nA (curve not shown), clearly indicating that this is one of the parameters controlling the nanogenerator performance. Please note that even for this thickness, our devices are within the usual range for NW-nanogenerators with ZnO single-crystalline nanowires length up to several micrometres.[10] An AC current was also obtained applying to the system a low frequency oscillator pressing at ~3 Hz the PDMS upper sheet generating the same periodical short-circuited piezo current (Figure 2 d-e)). It is possible to observe that the output signal quickly responds to the oscillator movement, with increasing piezo-current values for higher load (Figure 2 e); besides, the output signal is highly symmetric with similar values for positive and negative current spikes.

Finally, the maximum level of complexity will be achieved with the fabrication of 3D core@multishell piezoelectric nanogenerators. Following the advances in previous reports in the literature, we embedded the core@multishell nanowires in PMMA in order to build long lasting devices. This is a well-known approach to avoid concurrent short circuits in 1D nanodevices.[7,10,30] Besides, PMMA is easily etched by oxygen plasma allowing the electrical contact of the ZnO nanowires with the upper electrode. Both configurations, H2Pc@ZnO and H2Pc@Au@ZnO NWs were fabricated on ITO/PET substrates previously decorated with a ZnO thin film of 900 nm (see Figure 1 a-c) and then embedded in PMMA by spin coating. As mentioned above, the equivalent thickness for the ZnO shell was settled in 400 nm. Top view SEM image in Figure 1 i) shows the H2Pc@Au@ZnO system after a 15 minutes plasma etching treatment what partially reveals the surface of the longest NWs. The duration and/or power of the plasma source have to be adjusted in order to increase or decrease the exposed polymer-free ZnO area. The XRD diagram from the as grown ZnO and embedded NWs did not show any evidence of crystal degradation after the spin coating of the polymer and the subsequent plasma treatment. The I-V characteristics of these devices are remarkably symmetric for the positive and negative bias as shown in Figure 2 f). Besides, the figure also shows the inner gold shell increases the conductivity of the device. This agrees with an intimate and improved contact between the ZnO shell and the bottom electrode. Figure 2 g) shows the piezo-current obtained for the H2Pc@ZnO nanowires system. In this case, the maximum value reached is below 4 nA, a current smaller than for the 900 nm thin film NG indicating the system integrating the H2Pc@ZnO responds with a lower output that the solely seed layer. In fact, the shape of the curve is quite different, with shorter recovering times between positive and negative current spikes. Although the comparison with the thin film counterpart is not direct because the quantitative estimation of the active area of the contacted and embedded nanowires is not feasible, these results might indicate that the H2Pc single crystalline core plays an additional role apart from being a simple 1D template for the formation of the core@shell nanoarchitectures. Thus, these small-molecule nanowires depict a semiconducting behaviour, n- or p-type, depending on the particular characteristics of the corresponding organic or metalorganic precursor molecule. In this case, the H2Pc / ZnO system corresponds to a p – n junction that acts as internal rectifying interface being very likely one of the contributing factors responsible for the lower piezo-current observed. However, on the other hand, the short recovering time might be related to the high flexibility of the ONWs able to accommodate large deformations and mechanical stretching. At this point, it is worth stressing again that the plasma deposition of the ZnO shell does not alter the organic nanowire crystalline structure as elucidate from the HREM image in Figure S3, therefore we can assume that in this case, the system is working as an hybrid nanowire with flexible and elastic organic cores which may reduce the compression of the ZnO shell. Thus, the organic core decreases the device output power but, on the other hand, providing mechanical stability to the entire system. In a last step towards nanosystems integration and exploitation of the multifunctional character of the core@shell nanowires we built a nanogenerator based on the H2Pc@Au@ZnO nanowires. The results of this study are shown in Figure 2 h-i). In this case, we assume that the quantitative comparison with the H2Pc@ZnO nanowire device is possible providing that statistically both systems present the same ZnO piezoelectric shell. Thus, it is remarkable that these nanogenerators provide the highest output current obtained in our experiments with values under maximum load as high as 70 nA. Note that these values around two times higher than those corresponding to the 2.4 um ZnO device (Figure 2d-e), up to 5 times larger than for the equivalent 900 nm thin film NG (Figure 2b) and around 20 times larger than those corresponding to the H2Pc@ZnO nanowires (Figure 2g)). The Au inner layer intimately connects each ZnO grain in the ZnO shell with the conducting substrate providing a straightforward path for charging and discharging the system what given rise to the higher output current. The open circuit output voltage for the NWs nanogenerators were quite similar in both cases, with maximum values of ~170 mV for the nanowires (Figure 2 i), even higher than for the thin film counterpart. On the other hand, looking to the compressing-releasing curves (Figure 2 c) and g-h) it is evident that the H2Pc@Au@ZnO device resembles the shape of the ZnO thin film curves but with shorter recovering times. The 1D ONW template in the H2Pc@Au@ZnO system is providing mechanical stability to the system but with an overall decrease in flexibility provoked by the Au shell in comparison with the H2Pc@ZnO.

## 4. Conclusions

In this communication we have probed the versatility of the plasma-vacuum combined deposition for the synthesis of piezoelectric core@shell and core@multishell nanogenerators. Single-crystalline organic nanowires deposited in vacuum provides a 1D core for the plasma deposition of piezoelectric ZnO shells. The conformal plasma deposition on organic nanowires yields ZnO shells consisting in closely packed oriented nanocolumns that conformally coat the nanowires. The results indicate these ZnO oriented polycrystalline shell can compete in piezoelectric efficiency with the best results of the bibliography corresponding to single-crystalline ZnO structure. Besides, the tridimensional conformal nature of such polycrystalline ZnO shells are efficient to accommodate mechanical deformations thanks to the flexible and elastic nature of the organic cores. Thus, the organic nanowire scaffold adds flexibility and elasticity to the structure that can be bent repeatedly keeping their structural integrity and performance.

Apart from the industrial up-scalability, wafer scale compatibility, the absence of solvents and the energetic efficiency typically claimed as advantages of the plasma and vacuum deposition methodologies, the results clearly illustrate the possibilities of such methodologies for engineering the interfaces to improve the efficiency of the piezoelectric nanoarquitectures. Thus, the fabrication of a conformal and continuous Au nanoshell is proposed here as a way to develop an electrical connection between the piezoelectric shell and the electrodes. The results show a performace improvement in the measurement current values up to 20 times comparing with an equivalent ZnO shell.

The presented synthetic methodology can be expanded to the synthesis of 3D structures (i.e., nanotrees and multistacks) opening the way to future designs and optimizations of piezoelectric devices with complex and improved response and/or with the possibility of being integrated in energy converter devices.

**Acknowledgements**. We thank the AEI, MINECO (MAT2016-79866-R and MAT2013-42900-P) and the EU through cohesion fund and FEDER programs for financial support. VLF thanks the Talent HUB (MCSA) project. JRS-V thanks the University of Seville through the VI PPIT-US.

**Figures**

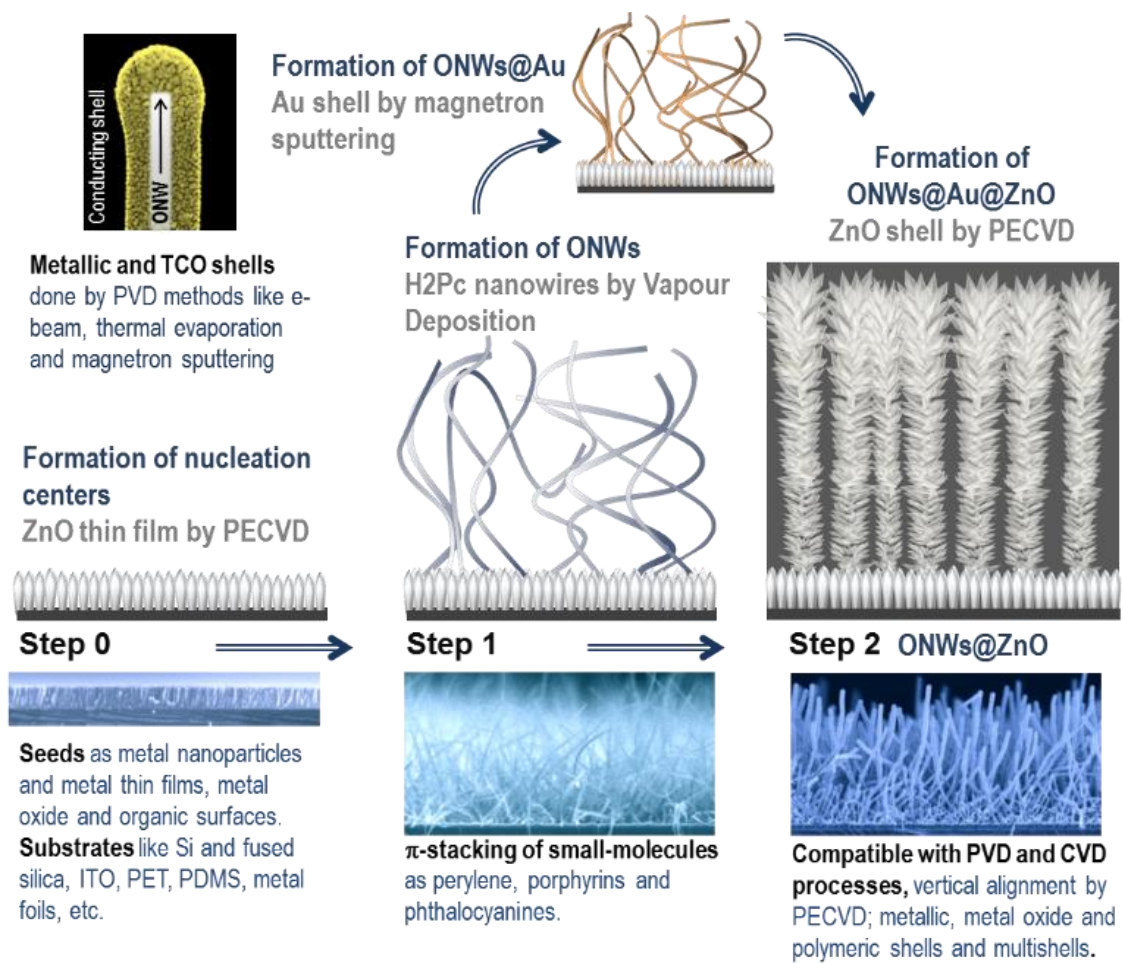

**Scheme 1.** Pictorial representation and representative SEM and STEM micrographs of the different steps involved in the formation of core@multishell supported nanowires.

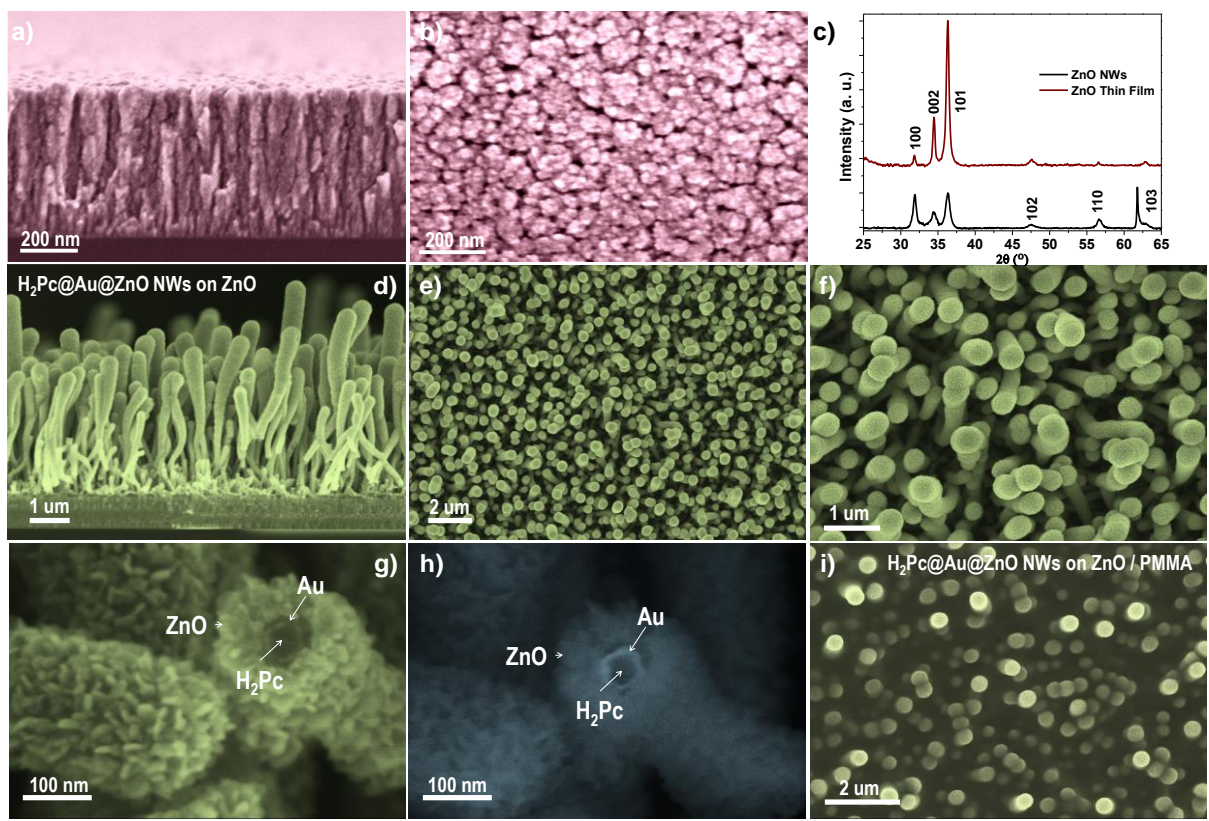

**Figure 1.** SEM micrographs showing the cross section (a) and planar view (b) of a PECVD ZnO thin film c) Ө-2Ө X-ray diffraction (XRD) patterns comparing the ZnO thin film with the Pc@ZnO nanowires; (d-f) Characteristics SEM images of the high density deposition of Pc@Au@ZnO NWs demonstrating the preferential vertical alignment of the 1D nanostructures after the plasma deposition of ZnO; g-h) these micrographs present correspondently the secondary electrons (g) and backscattered electrons images (h) micrographs of an individual nanowire revealing its internal composition; i) top view of the Pc@Au@ZnO NWs after being embedded in PMMA and subsequently treated with a oxygen plasma to reveal the nanocolumns tips.

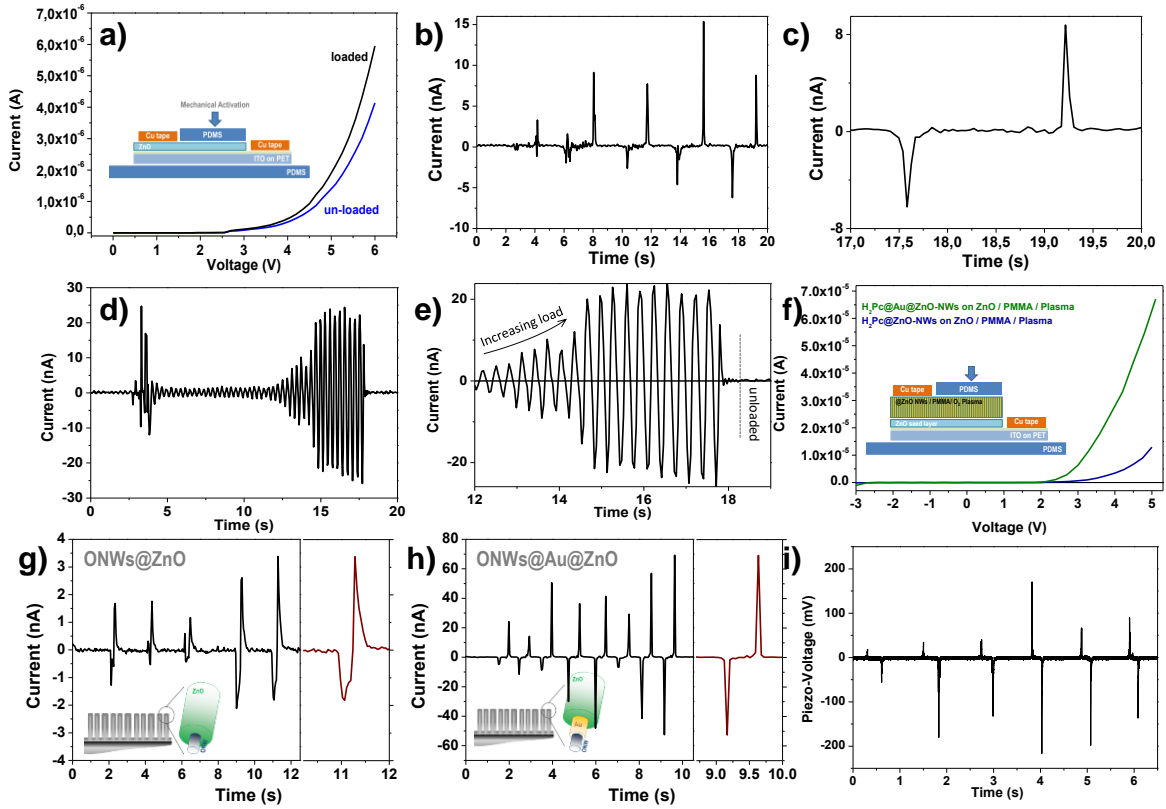

**Figure 2**. Characteristic I-V, I-t and V-t times obtained for the different devices; a) I-V curve for load and unload 900 nm ZnO thin film PENG, b-c) I-t short-circuited piezo-current output for intermittent deformation of the 900 nm ZnO thin film PENG; d-e) I-t short-circuited piezo-current output for intermittent deformation of the 2,4 um ZnO thin film PENG; f) I-V characteristics for the H2Pc@ZnO and H2Pc@Au@ZnO nanowires embedded in PMMA and contacted with Cu tapes after the oxygen plasma etching treatment; g-h) I-t short-circuited piezo-current output an expanded view of the marked regions for the @ZnO NW PENGs as labelled; i) Open-circuit voltage of the H2Pc@Au@ZnO PENG.

# SUPPORTING INFORMATION SECTION

## 3D core@multishell piezoelectric nanogenerators

A. Nicolas Filippin,[a] Juan R. Sanchez-Valencia,[a,b] Xabier García-Casas,[a] Víctor López-Flores,[a] Manuel Macias-Montero,[a] Fabian Frutos,[c] Angel Barranco[a] and Ana Borras[a]*

a) Nanotechnology on Surfaces Group, Materials Science Institute of Seville (CSIC-University of Seville), C/ Américo Vespucio 49, 41092, Seville, Spain.

b) Departamento de Física Atómica, Molecular y Nuclear, Universidad de Sevilla, Avda. Reina Mercedes s/n, 41012, Seville, Spain.

c) Departamento de Física Aplicada I, Universidad de Sevilla, Avda. Reina Mercedes s/n, 41012, Seville, Spain.

*Corresponding author e-mail: anaisabel.borras@icmse.csic.es

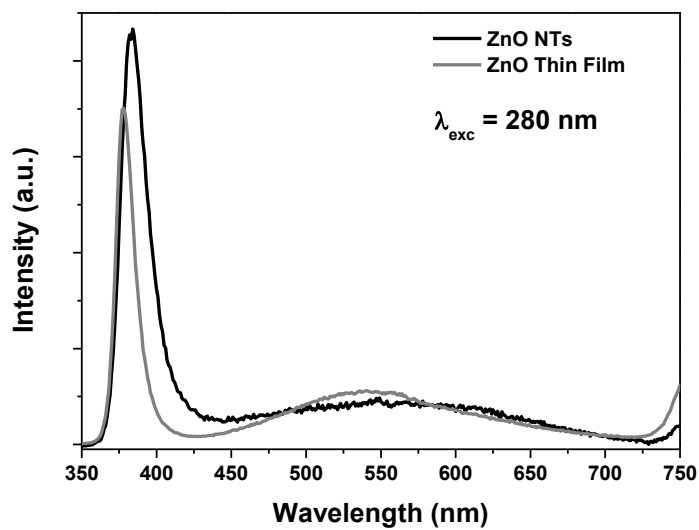

**Figure S1.** Room temperature photoluminescence emission spectra of the ZnO NTs and poly-crystalline thin film reference deposited in the same experimental conditions.

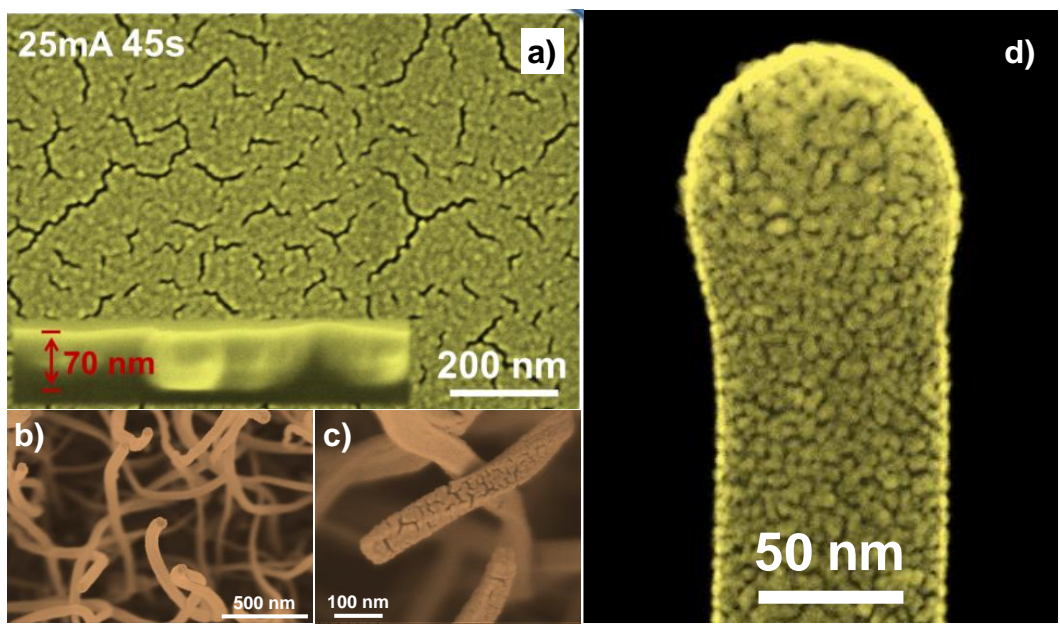

**Figure S2.** a-c) SEM and d) STEM micrographs showing the reference Au thin film (a) and conformal formation of the Au layer deposited by magnetron sputtering on the organic nanowires (b-d).

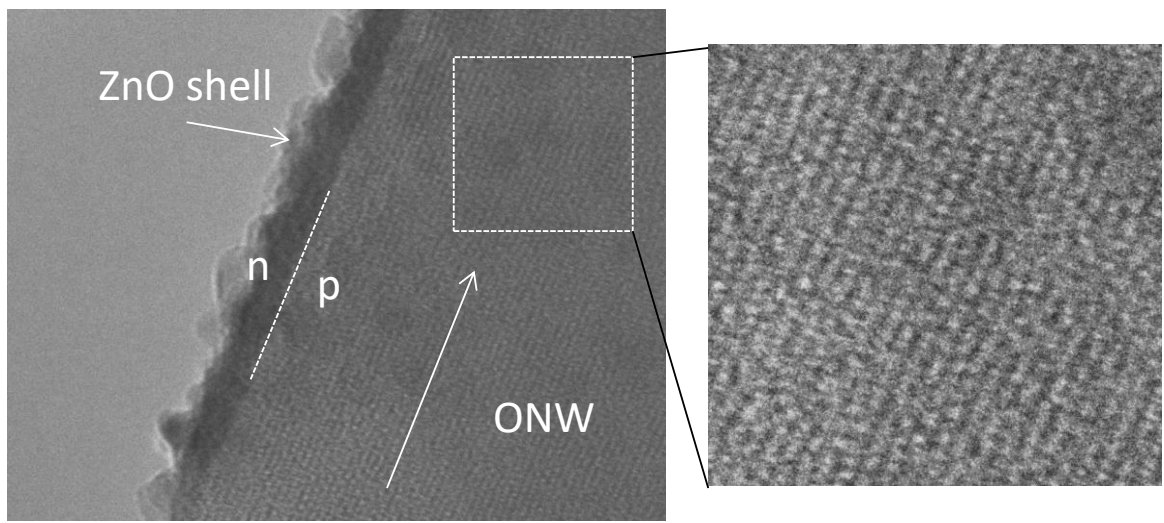

**Figure S3**. HREM micrograph of a single crystalline organic nanowire coated with a conformal ZnO shell.